\documentclass{osa-article}
\usepackage[title]{appendix}

\journal{oe}

\articletype{Research Article}

\begin{document}

\title{High-Contrast Integral Field Spectrograph (HCIFS): multi-spectral wavefront control and reduced-dimensional system identification}

\author{He Sun,\authormark{1,2,*} Alexei Goun,\authormark{3} Susan Redmond,\authormark{2} Michael Galvin,\authormark{2} Tyler Groff,\authormark{4} Maxime Rizzo,\authormark{4} and N. Jeremy Kasdin\authormark{2}}

\address{\authormark{1} Department of Computing and Mathematical Science, California Institute of Technology, Pasadena, CA 91125, USA}

\address{\authormark{2} Department of Mechanical and Aerospace Engineering, Princeton University, Princeton, NJ 08540, USA}

\address{\authormark{3} Department of Chemistry, Princeton University, Princeton, NJ 08540, USA}

\address{\authormark{4} NASA Goddard Space Flight Center, Greenbelt, MD, USA}

\email{\authormark{*}hesun@caltech.edu} 



\begin{abstract}
Any high-contrast imaging instrument in a future large space-based telescope will include an integral field spectrograph (IFS) for measuring broadband starlight residuals and characterizing the exoplanet's atmospheric spectrum. In this paper, we report the development of a high-contrast integral field spectrograph (HCIFS) at Princeton University and demonstrate its application in multi-spectral wavefront control. Moreover, we propose and experimentally validate a new reduced-dimensional system identification algorithm for an IFS imaging system, which improves the system's wavefront control speed, contrast and computational and data storage efficiency.
\end{abstract}

\section{Introduction}
\begin{figure}[h!]
\centering\includegraphics[width=13cm]{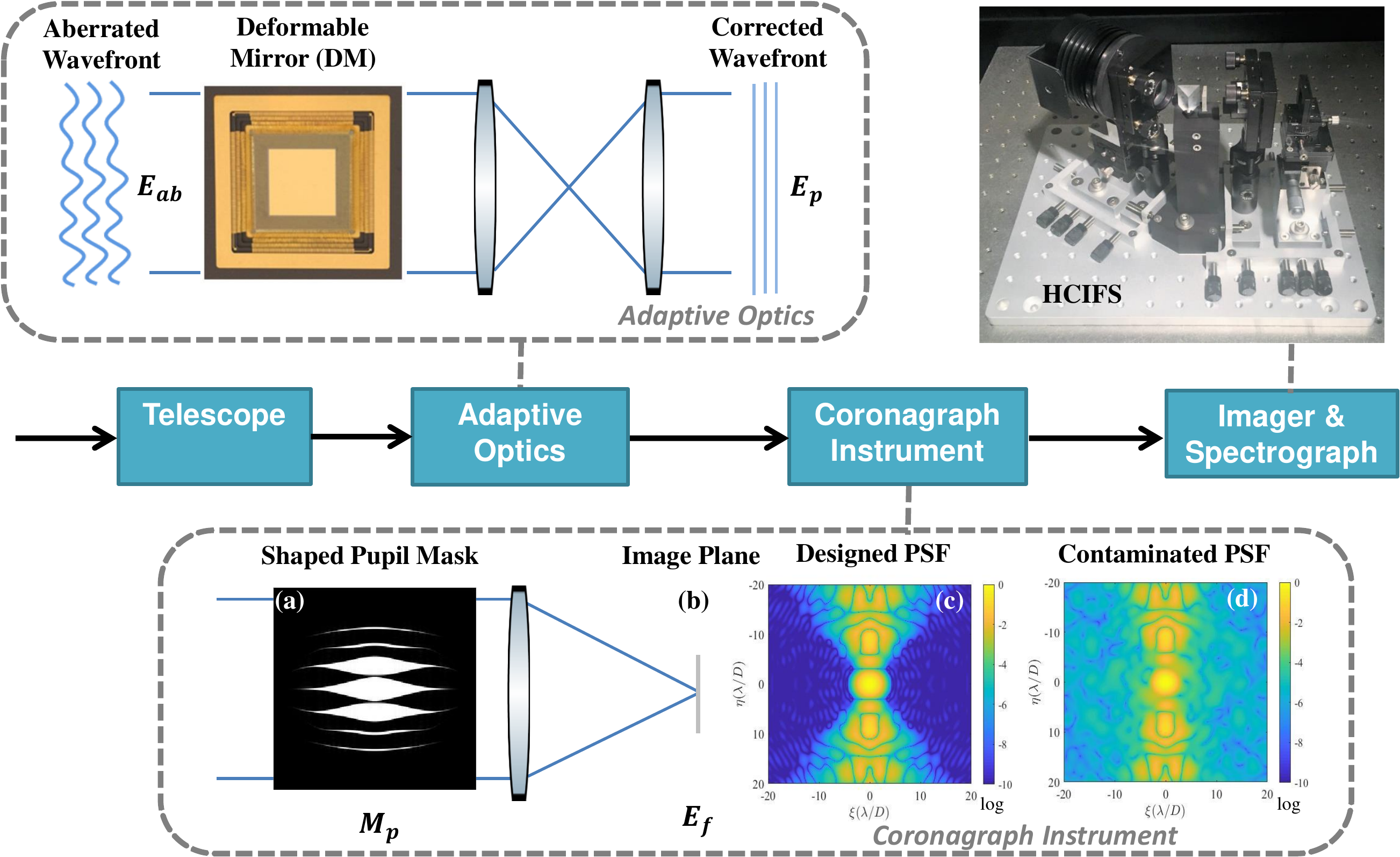}
\caption{Architecture of an example high-contrast instrument at Princeton's High-Contrast Imaging Laboratory. The coronagraph instrument uses (a) a shaped pupil mask\cite{kasdin2003extrasolar} to reshape the (b) point spread function (PSF) and create symmetric high-contrast regions (or so-called dark holes) in the image plane. However, the coronagraph is very sensitive to wavefront aberrations, which cause light leakage into the dark holes and decrease the contrast. The adaptive optics system corrects the complex wavefront aberrations (both phase and amplitude aberrations) using deformable mirrors\cite{bifano2011adaptive} to maintain the high contrast. The multi-spectral images are measured by an imager or a spectrograph. Here we show a photo of the high-contrast integral field spectrograph (HCIFS), which is a lenslet array-based IFS for simultaneously imaging and spectroscopy.}
\label{fig:instrument}
\end{figure}

Several high-contrast imaging instruments have been implemented in large ground-based telescopes, as well as being proposed for future space telescopes, for imaging faint exoplanets and characterizing their atmospheric compositions.\cite{ruane2018review, jovanovic2018review, snik2018review} A high-contrast imaging instrument, as shown in Fig.~\ref{fig:instrument}, mainly consists of a coronagraph and an adaptive optics (AO) system: the coronagraph\cite{kasdin2003extrasolar, guyon2003phase, mawet2009vector, zimmerman2016shaped, trauger2016hybrid} suppresses starlight that hides the planet signals, and the AO system\cite{macintosh2004extreme, verinaud2005adaptive, sun2019modern} corrects the residual starlight speckles caused by Earth's atmospheric turbulence or the telescope's optical aberrations. Typically, a high-contrast imaging instrument can achieve $10^{-3}$ contrast within the region of interest in ground-based telescopes\cite{guyon2010subaru, macintosh2014first, guyon2018extreme, currie2019no} and is predicted to allow $10^{-9}-10^{-10}$ contrast in future space telescopes\cite{mennesson2016habitable, bolcar2017large, seo2019testbed}. To study the chemical composition of a planet's atmosphere, a high-contrast instrument needs to operate in a wide bandwidth to obtain the planet's absorption spectrum. The broadband starlight field also needs to be measured for wavefront aberration corrections. An efficient approach to achieve both goals is integrating the high-contrast instrument with an integral field spectrograph (IFS).

An IFS is an optical instrument that combines spectrographic and imaging capabilities. Unlike a classical slit spectrograph, an IFS disperses the multi-spectral light in the entire field-of-view (FOV), so the entire broadband starlight field can be measured at once and the potential planet signals can be extracted without prior knowledge of their positions. An IFS has been widely implemented in current large ground-based telescopes, such as OSIRIS\cite{larkin2003osiris} for the Keck Telescope, GPI IFS\cite{chilcote2012performance} for the Gemini Telescope, CHARIS\cite{groff2015charis} for the Subaru Telescope, and has been preliminarily tested for space telescopes in lab, such as PISCES\cite{mcelwain2016pisces, groff2017wavefront} at NASA's Jet Propulsion Laboratory (JPL) Caltech. 

In this paper, we will present our recent development of a high-contrast IFS (HCIFS) at Princeton's High-Contrast Imaging Lab (HCIL), dedicated to prototyping the AO for future space-based high-contrast instruments. More specifically, we will focus on the IFS-based multi-spectral wavefront control with imperfect system modeling. Since multi-spectral wavefront control typically requires high-dimensional state-space modeling of the optical system, i.e., requiring large storage and many on-board computational resources, here we propose and experimentally validate a reduced-dimensional system identification method for adaptive multi-spectral wavefront control. The experimental results from HCIL have demonstrated that this method improves the multi-spectral wavefront control speed and the final contrast in a broad bandwidth, because it enables online model error correction. It also justifies the feasibility of using a reduced-dimensional model for adaptive optics.


\section{HCIFS: optical design and data cube extraction} \label{sec:hcifs}
In this section, we overview the optical design of HCIFS and explain the pipeline for retrieving monochromatic images from HCIFS's broadband measurements.

\begin{figure}[h!]
\centering\includegraphics[width=10cm]{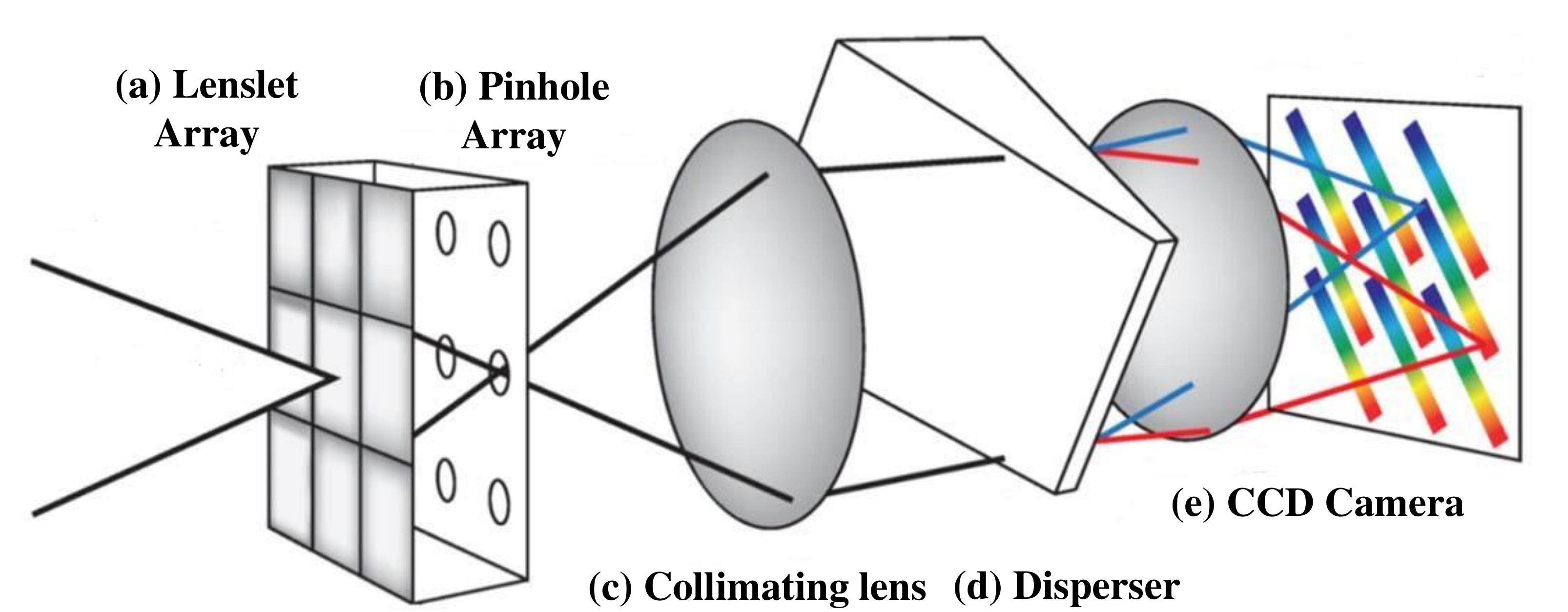}
\caption{Architecture of HCIFS by Rizzo et. al.\cite{rizzo2017simulating} The incident light is first down-sampled by (a) a lenslet array and (b) a pinhole array. Then the light is (c) collimated and sent to (d) a disperser (combination of prisms). The dispersed spectra are finally imaged on a CCD camera.}
\label{fig:ifs}
\end{figure}

As shown in Fig.~\ref{fig:ifs}, HCIFS is a lenslet array-based IFS. It mainly consists of three parts, (a)(b) a lenslet and pinhole array, (c)(d) a set of dispersion optics and (e) a CCD camera. The lenslet and pinhole array first down-samples the focal plane light field to a sparse field, then the prism disperses the light in the entire FOV and finally the CCD camera records the dispersed image. As a result, the 3-D (x, y and wavelength) broadband image is encoded as a 2-D spatio-spectral image on the detector. The initial down-sampling prevents crosstalk among spectra at different locations. Specifically, the lenslet and pinhole array of HCIFS is designed to sample the incident field at a Nyquist sampling rate, i.e., there  are two lenslets per $\lambda/D$ (wavelength-aperture size ratio) in one dimension. The FOV of HCIFS is $47 \times 39 \lambda/D$. The disperser has a two-component design (a Zinc Sulfide prism and a fused silica prism), which achieves a spectral resolution of 13.2 nm. HCIFS operates at 600 nm to 720 nm (120nm/660nm = $18\%$ bandwidth). More details of HCIFS's optical design are presented in references\cite{delacroix2018first, galvin2019rapid}.

\begin{figure}[h!]
\centering\includegraphics[width=13cm]{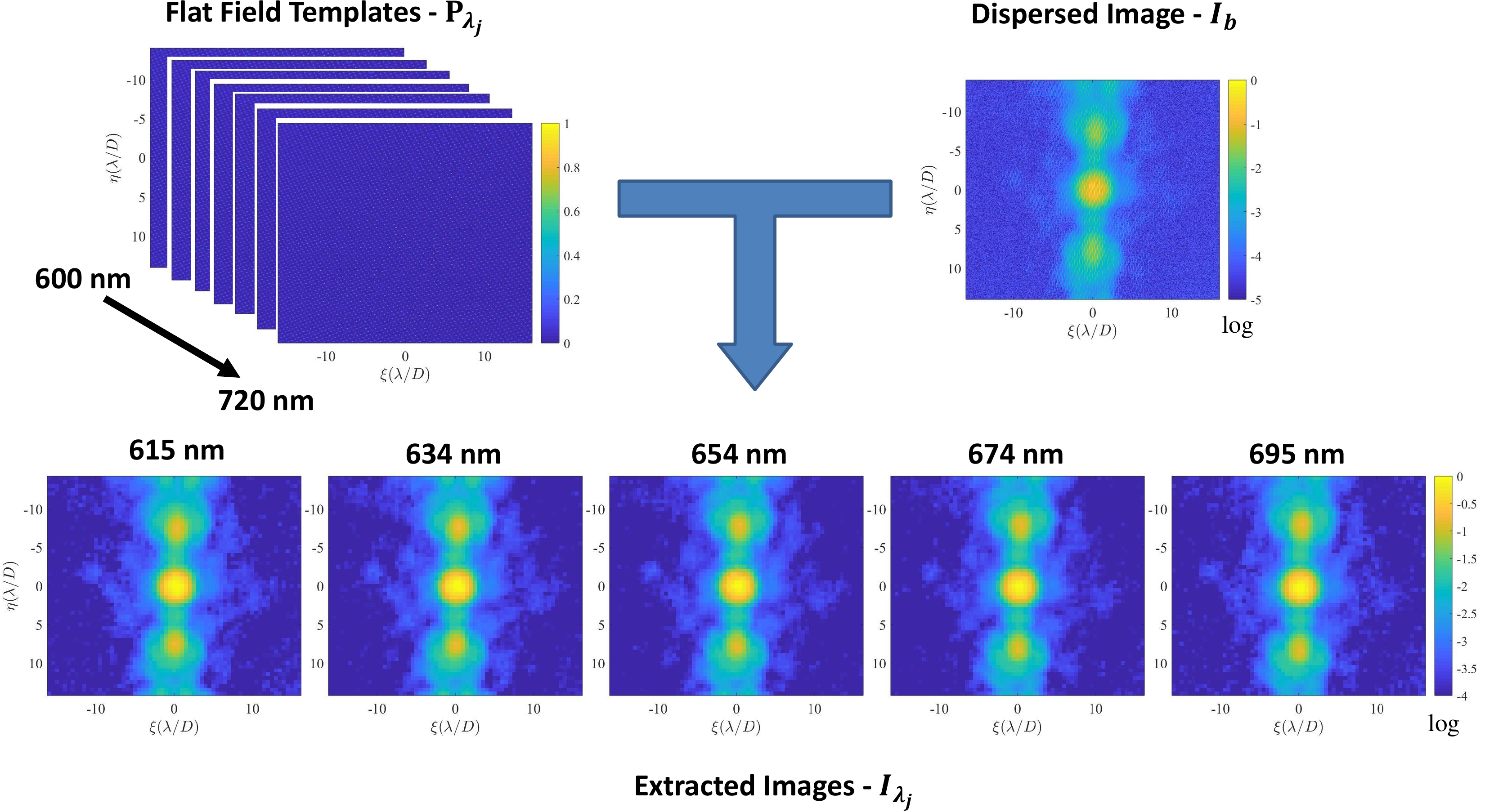}
\caption{HCIFS data cube extraction pipeline. Given pre-collected monochromatic flat field templates, a dispersed image can be translated to corresponding monochromatic images by solving a linear inverse problem. Here we show the contaminated dispersed PSF and reconstructed monochromatic PSF from HCIFS. $(\xi, \eta)$ are the focal plane coordinates. All images are scaled by $\lambda/D$, where $\lambda$ is the HCIFS's central operating wavelength (660 nm) instead of corresponding light wavelength and $D$ is the pupil size (1cm). The PSFs slightly become larger as the light wavelengths increase. It is not significant, but can be observed by checking the distances among speckles. For example, the distances between the top-right strip speckle and the PSF center increases from left to right.}
\label{fig:cube}
\end{figure}

The dispersed image is approximately a linear superposition of all monochromatic fields. Therefore, the 3-D image cube can be reconstructed by solving a linear inverse problem,
\begin{equation} \label{eq:extraction}
\min_{I_{\lambda_1}, \cdots, I_{\lambda_N}} \|I_b - \sum_{j=1}^N I_{\lambda_j} * P_{\lambda_j}\|_2^2,
\end{equation}
where $I_b$ is the HCIFS dispersed image, $\{I_{\lambda_j}\}$ are the reconstructed monochromatic images, N is the number of sampled wavelengths and $\{P_{\lambda_j}\}$ are the corresponding monochromatic flat field templates. This problem is typically referred to as ``data cube extraction"\cite{brandt2017data} in IFS image processing, which is a typical linear deconvolution problem. The flat field templates are measured by giving monochromatic input fields over the full bandwidth. Figure~\ref{fig:cube} shows a HCIFS data cube extraction result of a contaminated coronagraph PSF. Five slices of the reconstructed data cube are displayed. The monochromatic PSF slightly scales up as the wavelength increases, which agrees with Fourier optics that PSF size is proportional to light wavelength. The reconstructed image cube can be then utilized for the following broadband wavefront control and reduced-dimensional system identification.

\section{Methods: broadband control and reduced-dimensional system identification} \label{sec:method}
Wavefront control is a model-based stochastic control problem. Here we only consider correcting the instrumental wavefront aberrations in space-based AO, or so-called wavefront sensing and control (WFSC). In this case, the wavefront control only uses the image plane camera (or IFS), but no extra wavefront sensors for measuring the fast-evolving atmospheric turbulence.

The focal plane electric field of a high-contrast instrument can be represented as a linear state-space model\cite{groff2015methods}, as discussed in Appendix~\ref{sec:ssm},
\begin{equation} \label{eq:monossm}
    E_{f, k} = E_{f, k-1} + G \Delta u_k + w_k,
\end{equation}
where $E_{f, k} \in \mathbb{R}^{N_{pix} \times 1}$ is the focal plane field, $G \in \mathbb{R}^{N_{pix} \times N_{act}}$ is the Jacobian matrix, $\Delta u_k \in \mathbb{R}^{N_{act} \times 1}$ is the DM control voltage command, $w_k \in \mathbb{R}^{N_{pix} \times 1}$ is the additive Gaussian state transition noise, $k$ is the time step, $N_{pix}$ is the number of camera pixels in the dark hole and $N_{act}$ is the number of DM actuators. By concatenating the electric field, Jacobian matrices and state transition noises at different wavelengths, $E_{f, k}^{\lambda_j}$, $G_{\lambda_j}$ and $w_{f, k}^{\lambda_j}$, we can derive a state space model for the broadband electric field,
\begin{equation} \label{eq:broadbandssm}
    E_{f, k}^b = E_{f, k-1}^b + G_b \Delta u_k + w_k^b,
\end{equation}
where $E_{f, k}^b = [\cdots; E_{f, k}^{\lambda_j}; \cdots] \in \mathbb{R}^{(N \times N_{pix}) \times 1}$, $G_b = [\cdots; G_{\lambda_j}; \cdots] \in \mathbb{R}^{(N \times N_{pix}) \times N_{act}}$, $w_{k}^b = [\cdots; w_{k}^{\lambda_j}; \cdots] \in \mathbb{R}^{(N \times N_{pix}) \times 1}$ and N is the number of sampled wavelengths. The intensity of the electric fields are measured by the IFS reconstructed monochromatic images (Eq.~\ref{eq:extraction}),
\begin{equation} \label{eq:observation}
    I_{f, k}^b = [\cdots; I_{\lambda_j, k}; \cdots]= |E_{f, k}^b|^2 + I_{in, k}^b + n_k^b.
\end{equation}
where $I_{in, k}^b$ is the incoherent background, such as stray light or planet signals, and $n_k^b$ is the additive measurement noises, which is assumed to follow a Gaussian distribution but its covariance is proportional to the intensity (for approximating Poisson distribution). 

With the state-space model (Eq.~\ref{eq:broadbandssm}) and the observation model (Eq.~\ref{eq:observation}) defined, the wavefront control loop can be closed by first estimating the electric field based on the images, and then computing the DM command that removes starlight speckles and maintains a high contrast. 

Electric field estimation applies maximum likelihood estimation (MLE) methods, including batch process least squares regression\cite{borde2006high, give2011pair, sun2020efficient}, Kalman filtering\cite{groff2013kalman} or extended Kalman filtering\cite{riggs2016recursive, pogorelyuk2019dark}. It collects measurements by introducing various DM sensing commands, and solves for the hidden electric field by minimizing the difference between measurements and model predictions as well as constraining the solution close to the electric field of the last step. For example, the extend Kalman filter computes the hidden electric field via
\begin{equation} \label{eq:estimation}
    \begin{split}
        &\min_{E_{f, k}^b, I_{in, k}^b} \ \sum_{m=1}^{N_{obs}}\|\bar{I}_{f, k}^{b, m} - |\bar{E}_{f, k}^{b, m}|^2-I_{in, k}^b\|_{R_k^{-1}}^2 + \|E_{f, k}^b - \bar{E}_{f, k}^b\|_{Q_k^{-1}}^2\\
        & \quad s.t. \quad \quad \bar{E}_{f, k}^{b, m} = E_{f, k}^b + G_b \Delta \bar{u}_k^m, \quad \forall m = 1, \cdots, N_{obs}\\
        & \quad \quad \quad \quad \ \bar{E}_{f, k}^b = E_{f, k-1}^b + G_b \Delta u_k,
    \end{split}
\end{equation}
where $\bar{I}_{f, k}^{b, m}$ and $\bar{E}_{f, k}^{b, m}$ are, respectively, the sensing images and sensing fields, $\bar{E}_{f, k}^b$ is the electric field prediction based on last step, $\Delta \bar{u}_k^m$ are the DM sensing commands, $N_{obs}$ is the number of sensing images, and $\|\cdot\|_{R_k^{-1}}$ and $\|\cdot\|_{Q_k^{-1}}$ are the two weighted norms ($Q_k$ and $R_k$ are respectively the covariance matrices of the Gaussian transition noises, $w_k^b$, and the Gaussian measurement noises, $n_k^b$, at the $k$-th time step; a weighted norm defined by matrix $W$ is $\|x\|_W = x^T W x$), which balance the measurement term and the prior knowledge term.

Wavefront controllers then use the estimated electric field to find the optimal control command. Common wavefront controllers include electric field conjugation (EFC)\cite{borde2006high, give2007broadband}, stroke minimization (SM)\cite{pueyo2009optimal} and robust linear programming controller (RLPC)\cite{sun2017improved}. For example, EFC is a quadratic controller that minimizes the starlight speckle intensity,
\begin{equation} \label{eq:control}
    \min_{\Delta u_{k+1}} \|E_{f, k}^b - G_b \Delta u_{k+1}\|_2^2 + \alpha_k \|\Delta u_{k+1}\|_2^2,
\end{equation}
where $\alpha_k$ is the Tikhonov regularization parameter that prevents unreasonably large control command. Looping the above estimation and control steps, the contrast is gradually improved and finally maintained at a high level for scientific observations.

Both the electric field estimation and the wavefront control require an accurate Jacobian matrix (Eq.~\ref{eq:broadbandssm}). However, in a real instrument, manufacturing errors, mis-calibrations, thermal effects and modeling approximations (e.g. Fourier optics approximation) always cause mismatches between the real system and the state-space model. Recently, data-driven approaches have been developed to achieve system identification using a telescope's point spread function (PSF) images \cite{doelman2019identification} or real-time wavefront control data \cite{sun2018identification, sun2018neural}. One of the experimentally verified approaches \cite{sun2018identification} uses an E-M algorithm to identify model parameters, which iteratively reconstructs the hidden electric field (E-step) and updates system's Jacobian matrix (M-step). 
The E-step is identical to the electric field estimation as in  Eq.~\ref{eq:estimation}, where the model is assumed known and the hidden electric field is estimated; the M-step, to the contrary, minimizes the same cost function where we assume the hidden electric field is known but the model parameters, i.e., the Jacobian matrix, are optimized. However, since the electric field estimation from the E-step is a probability distribution instead of a point estimate, the M-step is a stochastic optimization problem instead of a deterministic optimization problem (constraints are omitted because they are same as those in Eq.~\ref{eq:estimation})
\begin{equation} \label{eq:mstep}
\min_{G_b} \ \sum_{m=1}^{N_{obs}} \langle \|\bar{I}_{f, k}^{b, m} - |\bar{E}_{f, k}^{b, m}|^2-I_{in, k}^b\|_{R_k^{-1}}^2 + \|E_{f, k}^b - \bar{E}_{f, k}^b\|_{Q_k^{-1}}^2 \rangle,
\end{equation}
where $\langle \cdot \rangle$ represents the expectation of the cost function given $E_{f, k}^b$ and $I_{in, k}^b$ follow the estimated distributions from the E-step. The M-step can be solved using a stochastic gradient descent (SGD) method, where the Jacobian matrix is initialized based on the Fourier optics model. This approach has been demonstrated in experiment for monochromatic wavefront control.\cite{sun2018identification, sun2018neural}

In this paper, we adapt this algorithm to the IFS-integrated multi-spectral high-contrast imaging instrument. In such a system, a high-dimensional state-space model is required as the system controls the wavefront of multiple wavelengths. Therefore we propose to improve the computational efficiency by combining the current system identification method with a model reduction technique. Instead of using a full rank Jacobian matrix, we represent the Jacobian matrix as $G_b = U V^T$, where $U \in \mathbb{R}^{(N \times N_{pix}) \times r}$, $V \in \mathbb{R}^{N_{act} \times r}$ are lower ranked matrices (rank $r < \min \{N \times N_{pix}, N_{act}\}$, which physically means the number of modes in the reduced-dimensional Jacobian matrix). This low rank approximation is expected to save model parameters by capturing the correlations of different pixels and different actuators, as well as the similarity among electric fields at different wavelengths. Therefore, the M-step becomes,
\begin{equation} \label{eq:reduced}
    \begin{split}
        &\min_{U, V} \ \sum_{m=1}^{N_{obs}} \langle \|\bar{I}_{f, k}^{b, m} - |\bar{E}_{f, k}^{b, m}|^2-I_{in, k}^b\|_{R_k^{-1}}^2 + \|E_{f, k}^b - \bar{E}_{f, k}^b\|_{Q_k^{-1}}^2 \rangle\\
        & \quad s.t. \quad \quad \bar{E}_{f, k}^{b, m} = E_{f, k}^b + U V^T \Delta \bar{u}_k^m, \quad \forall m = 1, \cdots, N_{obs}\\
        & \quad \quad \quad \quad \ \bar{E}_{f, k}^b = E_{f, k-1}^b + U V^T \Delta u_k,
    \end{split}
\end{equation}
which can be also solved using SGD method. This would be extremely beneficial for future large space telescopes that have DMs with more than ten thousand actuators and aim for large high-contrast observation areas in the image plane. The low rank assumption reduces the computation and data storage complexity of the M-step to $\text{O} (r \times (N \times N_{pix} + N_{act}))$ from $\text{O}(N \times N_{pix} \times N_{act})$ for the standard E-M algorithm. Also, the low-rank assumption is an implicit regularizer during system identification, which prevents the identified Jacobian matrix from overfitting to the measurement data noises. We experimentally test this reduced-dimensional system identification and the resulting wavefront control in Sec.~\ref{sec:result}.

\section{Results and discussion} \label{sec:result}
\begin{figure}[h!]
\centering\includegraphics[width=13cm]{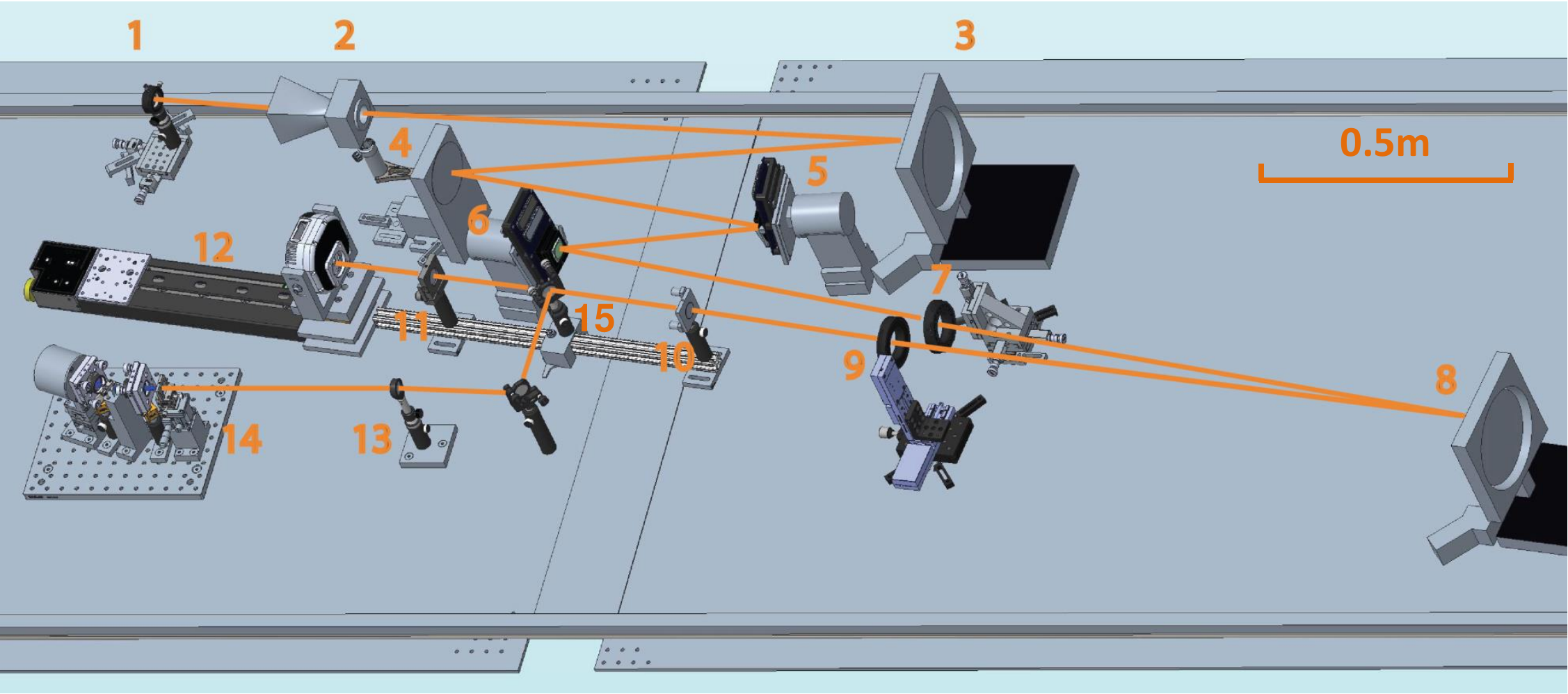}
\caption{Layout of the Princeton’s High Contrast Imaging Laboratory (HCIL).\cite{delacroix2018first} The list of devices: 1.~laser, 2.~baffle, 3.~first off-axis parabola (OAP1), 4.~fold mirror, 5.~DM1, 6.~DM2, 7.~shaped pupil mask (SP), 8.~second off-axis parabola (OAP2), 9.~focal plane mask (FPM), 10~and~11.~reimaging lenses, 12.~CCD camera, 13.~focusing lens, 14.~integral field spectrograph (IFS), 15.~pick-off mirror.}
\label{fig:hcil}
\end{figure}
\begin{figure}[h!]
\centering\includegraphics[width=11.5cm]{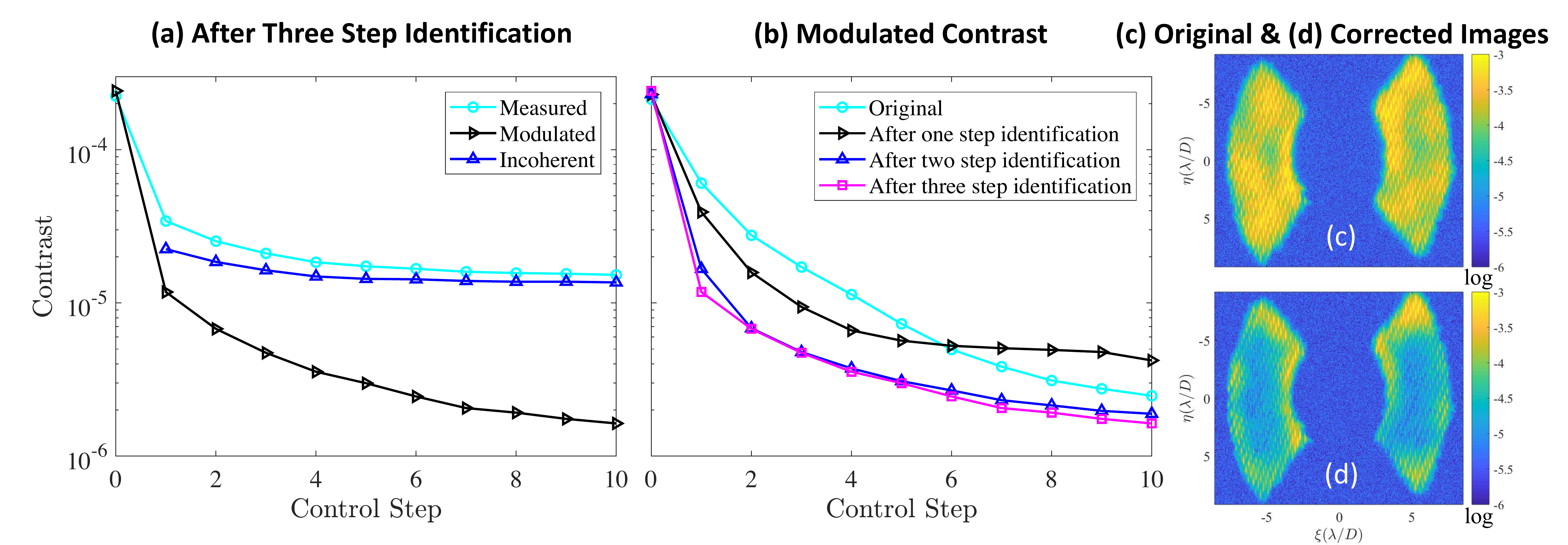}
\caption{Multi-spectral wavefront control with reduced-dimensional system identification. The rank of the reduced Jacobian matrix is 400. (a)~Measured, modulated and incoherent contrast versus control step after three system identification trials. (b)~Modulated contrast versus control step after each system identification trial. (c)~Original (before adaptive optics correction) and (d)~corrected HCIFS dispersed images. As can be seen, the wavefront control becomes faster and achieves higher contrast after system identification. The reduced-dimensional model performs robust wavefront control.}
\label{fig:control}
\end{figure}

The experiments are conducted using  Princeton's HCIL testbed, whose layout is shown in Fig.~\ref{fig:hcil}. It utilizes a shaped pupil mask (as shown in Fig.~\ref{fig:instrument} (a)) to create high-contrast observation regions. In addition, a bowtie-shaped focal plane mask is used to block the central bright part of the PSF to avoid camera saturation.  This only allows the light in the dark zone crescents on either side of the main PSF to propagate to the detector as shown in Fig.~\ref{fig:control}~(c)(d)). A pair of continuous surface MEMS DMs
from Boston Micro-machines Corporation (BMC)\cite{bifano2011adaptive} are used to correct the wavefront aberrations in the system. Each DM has 952 actuators. A Koheras SuperK Compact super-continuum laser source (equipped with
eleven narrow-band filters and one broadband filter) is used for simulating the multi-spectral starlight. We also have two detectors on the testbed, a CCD camera and the HCIFS, between which we can switch using a pick-off mirror. In all our experiments, we use Python package ``CRISPY"\cite{rizzo2017simulating} to extract the monochromatic images from IFS measurements. The broadband state is defined by an electric field at five representative wavelengths, 615nm, 634nm, 654nm, 674nm, 695nm. Our control law tries to improve the contrast in two symmetric sectors of the images (as shown in Fig.~\ref{fig:control}) with a 60 degree angular range and a 3 $\lambda/D$ width (6 to 9 $\lambda/D$ from the center).   

Figure~\ref{fig:control} shows the wavefront control and system identification results where we only keep 400 modes of the Jacobian matrix, i.e. the rank of matrices $U$ and $V$ are constrained to be 400 ($r=400$). The test was run as follows: we run wavefront control for ten loops with a fixed Jacobian matrix, then we update the Jacobian matrix using the reduced-dimensional E-M algorithm and run another ten steps of wavefront control using the updated model, repeating until the wavefront correction no longer improves (not faster or achieving higher contrast). Figure~\ref{fig:control}~(a) reports the contrast-versus-control step curve after the model accuracy converges. The three lines respectively represent the modulated contrast ($|E_{f, k}^b|^2$, contrast of starlight which can be influenced via wavefront control), the incoherent contrast ($I_{in, k}^b$, contrast of stray light or incoherent planet signals not influenced by wavefront control), and the broadband measured contrast ($I_{f, k}^b$, summation of the modulated contrast and the incoherent contrast). All three contrasts are computed by averaging over the five representative wavelengths. Currently, our IFS-based high-contrast imaging instrument is mainly limited by the incoherent light (mainly from the laser fiber). 
Figure~\ref{fig:control}~(b) shows the modulated contrast curve after each system identification trial. With system identification, the wavefront control reaches a high contrast with fewer control steps and achieves a higher final contrast compared with the experiment using the original biased model. The correction speed and contrast stop improving after around three system identification trials.

\begin{figure}[h!]
\centering\includegraphics[width=13cm]{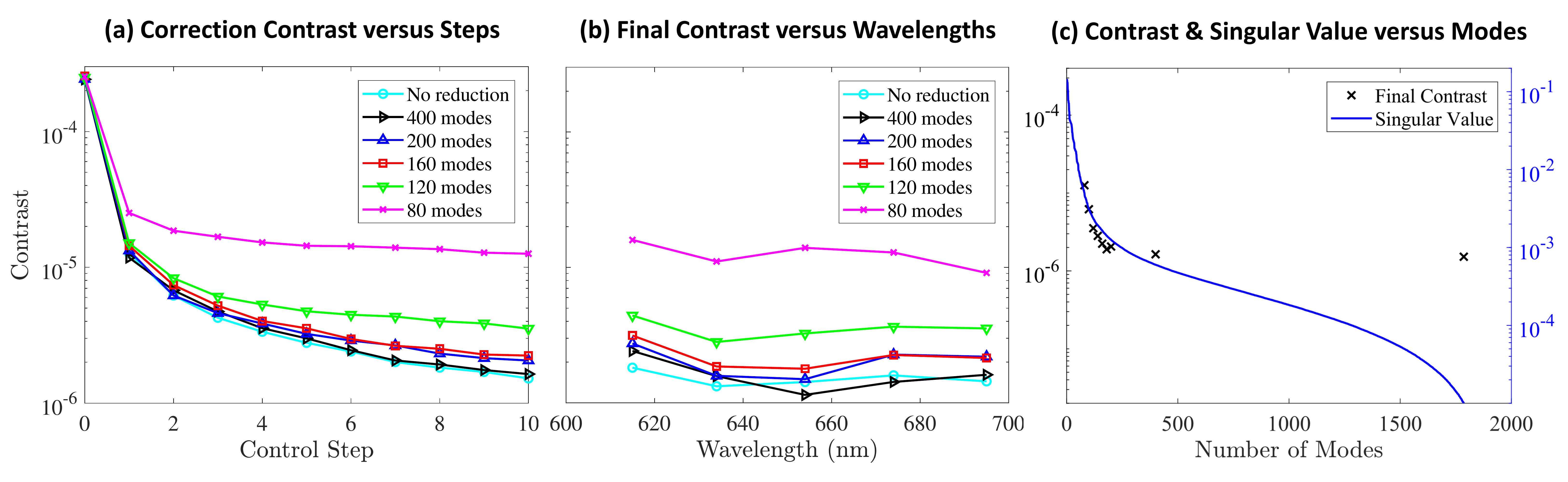}
\caption{System Identification with different numbers of modes (rank of identified Jacobian matrix). (a) Contrast versus control step. (b) Contrast over bandwidth. (c) Comparison of the final achievable contrast and singular values. The contrast is approximately linearly related to the singular value at the specific number of modes, so the reduced model's rank can be determined based on the desired system's contrast.}
\label{fig:mode}
\end{figure}

In Fig.~\ref{fig:mode}, we report the wavefront control results after three system identification trials with different numbers of modes. Reducing the number of modes influences the final contrast of the system. However, the contrast does not improve significantly after having more than 200 modes, which indicates that 200 modes should be enough for approximating our control system. Figure~\ref{fig:mode}~(c) compares the final contrast and the singular values of the full-rank Jacobian matrix. As can be seen, the contrast is approximately proportional to the cut-off singular value, so a reasonable number of modes for the Jacobian matrix can be selected according to the instrument's target contrast. The final contrast stops improving with more than 400 modes. This is likely because our system's achievable contrast is limited by the bright stray light, since the accuracy of electric field estimation will be highly influenced by the stray light photon noise. Our future work is focusing on reducing the stray light in the instrument, such as replacing the laser fiber and introducing a spatial filter.

\section{Summary} \label{sec:summary}
In this paper, we have reported our development of a high-contrast integral field spectrograph (HCIFS) for a telescope's high-contrast imaging instrument and its multi-spectral wavefront control approach and results. Moreover, we propose a reduced-dimensional system identification method for improving the instrument's modeling accuracy. Experimental results demonstrate that the identified reduced-dimensional model improves the system's wavefront control speed, final contrast and computation efficiency. The code for reduced-dimensional system identification has been published on Github: https://github.com/HeSunPU/ML4AO.

In future work, we plan to extend this reduced-dimensional system identification method to more complicated adaptive optics systems. For example, in high frequency wavefront sensor based adaptive optics system\cite{frazin2013utilization}, the method will allow us to characterize not only the errors in the main optical path, but also the biases in wavefront sensor measurements.
\begin{appendices}
\section{State-space modeling of a high-contrast imaging instrument} \label{sec:ssm}
As shown in Fig.~\ref{fig:instrument}, with the aberrated incident wavefront, the DM surface deformation, the corrected pupil plane wavefront, the coronagraph mask, the image planet field denoted as $E_{ab}$, $\phi$, $E_p$, $M_p$ and $E_f$ respectively, we have
\begin{equation} \label{eq:optical_model}
    \begin{split}
        E_p &= E_{ab} \exp(i \frac{4 \pi \phi}{\lambda}),\\
        E_f &= \mathcal{F}\{M_p E_p\},
    \end{split}
\end{equation}
where $\mathcal{F}\{\cdot\}$ is a Fourier transform operator. The mirror deformation is approximately a linear superposition of each actuator’s influence,
\begin{equation} \label{eq:inf}
    \phi = \sum_{q=1}^{N_{act}} u_q f_q, 
\end{equation}
where $N_{act}$ is the number of actuators on the DM, $q$ is the actuator index, $u_q$ is the DM voltage command, and $f_q$ is a single actuator's influence on the DM surface deformation (referred to as the DM influence function). Writing the DM voltage command as a time-cumulative formula,
\begin{equation} \label{eq:cum}
    u_{q} = u_{q, k} = u_{q, k-1} + \Delta u_{q, k},
\end{equation}
we can derive a linear state transition model (combining Eq.~\ref{eq:optical_model}, Eq.~\ref{eq:inf} and Eq.~\ref{eq:cum}) assuming the DM adjustment at each step is small (typically even accumulated DM adjustments are only a few tenths of wavelength),
\begin{equation} \label{eq:linearization}
\begin{split}
    E_{f, k} &= \mathcal{F}\{E_{ab} M_p \exp(i \frac{4 \pi \sum_{q=1}^{N_{act}} (u_{q, k-1} + \Delta u_{q, k}) f_q}{\lambda})\} \\
    &\approx \mathcal{F}\{E_{ab} M_p \exp(i \frac{4 \pi \sum_{q=1}^{N_{act}} u_{q, k-1} f_q}{\lambda}) [1 + i \frac{4 \pi \sum_{q=1}^{N_{act}} \Delta u_{q, k} f_q}{\lambda}]\}\\
    & = E_{f, k-1} + \sum_{q=1}^{N_{act}} \frac{i 4 \pi E_{f, k-1} \mathcal{F}\{f_q\}}{\lambda} \Delta u_{q, k}\\
    &\approx E_{f, k-1} + \sum_{q=1}^{N_{act}} \frac{i 4 \pi \mathcal{F}\{E_{ab} M_p f_q\}}{\lambda} \Delta u_{q, k}.
\end{split}
\end{equation}
This can be denoted as a  standard state-space formula (neglecting noises),
\begin{equation} \label{eq:ssm}
    E_{f, k} = E_{f, k-1} + G \Delta u_k,
\end{equation}
where $\Delta u_k = [\Delta u_{1, k}, \cdots, \Delta u_{N_{act}, k}]$ and $G$ is the Jacobian matrix computed from $E_{ab}$, $M_p$, $f_q$ and $\lambda$. Electric fields at different wavelengths have different state space response. 

\end{appendices}

\section*{Funding}
This work is supported by NASA Grant No. 80NSSC17K0697 (GSFC). 

\section*{Acknowledgments}
The authors would like to thank Mary Anne Limbach for the discussions on IFS design.

\section*{Disclosures}
The authors declare no conflicts of interest.

\bibliography{sample}






\end{document}